\documentclass[11pt,journal,draftclsnofoot,a4paper,onecolumn]{IEEEtran}
\usepackage{graphicx, amsmath, amssymb}
\usepackage{enumerate}
\usepackage{breqn}
\usepackage[scriptsize]{subfigure}
\usepackage{color}
\usepackage{float}
\usepackage{graphicx, amsmath, amssymb}
\usepackage{cases}
\usepackage[justification=centering]{caption} 
\usepackage{enumerate}
\usepackage[scriptsize]{subfigure}
\usepackage{graphicx}
\usepackage{subfigure}
\usepackage{algorithm}
\usepackage{algorithmic}        %用到的宏包，要自己改下
\usepackage{multirow}
\usepackage{leftidx}
\usepackage{graphicx}
\usepackage{subfigure}
\usepackage{float}

\newcommand{\abs}[1]{\left\vert#1\right\vert}

\DeclareMathOperator{\erf}{erf}

\graphicspath{{images/}}
\IEEEoverridecommandlockouts
\begin{document}
\title{Zero-Delay Gaussian Joint Source-Channel Coding for the Interference Channel}
\author{Xuechen Chen\\School of Electronics and Information Technology\\Sun Yat-Sen University, Guangzhou, China \\ chenxch8@mail.sysu.edu.cn}
\maketitle
\begin{abstract}
This paper studies zero-delay joint source channel coding (JSCC) for transmission of correlated Gaussian sources over a Gaussian interference channel (GIC). We propose to adopt delay-free hybrid digital and analog (HDA) scheme, which is, transmitting the superposition of scaled source and its quantized version after applying scalar quantization to the source at each transmitter. At the corresponding receiver, two kinds of estimators are presented. It is shown that both the schemes, when optimized, beat the uncoded transmission if the channel signal-to-noise ratio (CSNR) is higher than a threshold value for different correlation coefficients and interference values. 
\end{abstract}
\section{Introduction}
In many emerging applications involving wireless sensor networks (WSN), low-cost sensors with limited processing capabilities and battery life are deployed which implies that encoders with low complexity are needed. Such applications usually require real-time monitoring and control of underlying physical systems, which impose strict delay constraints. As a result, novel coding methods instead of traditional long block codes for separate source and channel coding (SSCC) which exhibit high complexity and high delay are needed.
 
We consider the extreme of \emph{zero-delay} problem, i.e., the transmission is to be done in a \emph{scalar} fashion. It is well known that a zero-delay uncoded (i.e., scale-and-transmit) scheme can achieve the minimum squared distortion for a Gaussian source transmitted over an additive white Gaussian noise (AWGN) channel with an input power constraint. However, this is not the case for multi-terminal problems, in general. Also, in multi-terminal scenarios, the optimality of SSCC breaks down. Note that in WSN, measurements collected by the sensors close to each other exhibit statistical correlation. As the correlation of sources can be adopted to generate correlated channel inputs by JSCC, it is attractive to turn to JSCC. As a matter of fact, uncoded scheme is a special case of JSCC. In recent years, many works have been done to understand multi-user JSCC problems. For example, \cite{Lapidoth} derived the distortion lower bound when a bivariate Gaussian is transmitted over a Gaussian multiple access channel (GMAC). The necessary conditions for optimality of uncoded transmission for multi-user communications over Gaussian broadcast channel (BC) and GMAC were generalized in \cite{Tian}. \cite{Floor} proposed two distributed delay-free JSCC schemes for a bivariate Gaussian on a GMAC. \cite{Xue, Varasteh} investigated zero-delay JSCC problems for a Gaussian source in the Wyner-Ziv Setting and over dirty-tape channel respectively.

In this paper, we consider zero-delay transmission of a pair of correlated Gaussian sources $(S_1,S_2)$ over a two-user Gaussian interference channel (IC). Each of two separate transmitters observes a different component of the source pair and describes the observations to the corresponding destination over a Gaussian IC. Receivers $i$ tries to recover the source $S_i$, where $i\in\{1,2\}$ with the minimum average distortion.  \cite{W.Liu} and \cite{P.Minero} gave achievable distortion region for JSCC IC problem in the lossless setting and lossy setup respectively. About JSCC Gaussian IC problem, \cite{denial1} derived an outer bound on the achievable region when the interference is weak and showed the condition for optimality of uncoded transmission. For strong interference case, see \cite{denial2}. Our goal here is to design low delay and low complexity JSCC techniques motivated by hybrid digital and analog strategies proposed in \cite{P.Minero}. After applying scalar quantizer to each source, the source itself and the quantized value are scaled and superimposed to be channel inputs. At each receiver, we propose two reconstruction methods. We present numerical results to show that as long as CSNR is higher than a threshold value, the proposed schemes offer better performance than uncoded transmission for different correlation coefficients and interference values.

     The remainder of the letter is organized as follows. Section II introduces the GIC problem while our scalar HDA encoding method and two kinds of estimation schemes are presented in Section III. Simulation results are given in Section IV. The supplementary file for this paper can be found in \cite{xue}.

\begin{figure}[ht]
\centering
\includegraphics[height = 1.6cm]{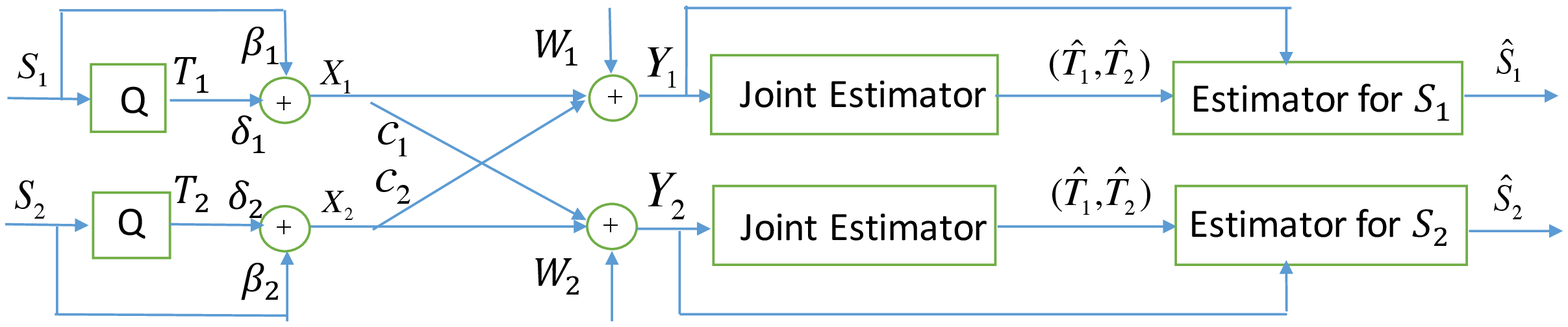}
\caption{Framework of Our Schemes}
\label{figr:framework}
\end{figure}

\vspace{-0.4cm}
\section{Problem Statement} 
We assume that a sequence of zero mean bivariate Gaussian source $\{S_{1,j},S_{2,j}\}_{j=1}^{\infty}$ is independent and identically distributed (i.i.d.) along time $j$, and the covariance matrix for each $(S_1,S_2)$ is
\[
\left[
\begin{array}{cc}
1 & \rho \\
\rho & 1
\end{array}
\right] \; 
\]
with $\abs{\rho}<1$. In other words, $S_i\sim{\mathcal N}(0,1)$ and
\[S_{i^c} = \rho S_i + N_i\quad\quad i=1,2,\quad i^c=\{1,2\}\backslash i\]
with $N_i\sim{\mathcal N}(0, \sigma^2_N)$, where $\sigma_N^2=1-\rho^2$ and $N_i$ is not only independent of $S_{i}$, but also of $N_{i^c}$. 

At the $i$-th transmitter, an $n$-block of the $i$-th source $\{S_{i,j}\}_{j=1}^n$ is to be mapped to channel input $\{X_{i,j}\}_{j=1}^n$ which should satisfy individual average power constraints,
\begin{align*}
\frac{1}{n}\sum\limits_{j=1}^n\mathbb{E}[|X_{i,j}|^2]\leq P_i,\quad\quad i = 1,2.
\end{align*}

Each $X_i$ then goes through an additive Gaussian IC with i.i.d. noise $W_i\sim{\mathcal N}(0,\sigma_W^2)$ whose output is $Y_i$, i.e., 
\begin{align*}
Y_i = X_i+c_{i^c}X_{i^c}+W_i\quad\quad i=1,2,
\end{align*}

The source is recovered to be $\{\hat{S}_{i,j}\}^n_{j=1}$ as a function of $\{Y_{i,j}\}_{j=1}^n$. The quality of the reconstruction is measured by the mean-squared-error distortion, i.e., 
\begin{align*}
D_i= \frac{1}{n}\sum\limits_{j=1}^n\mathbb{E}\Big[\big(S_{i,j}-\hat{S}_{i,j}\big)^2\Big].
\end{align*}
Then $D=\frac{1}{2}(D_1+D_2)$ denotes \emph{average} end-to-end distortion.
 
At the extreme of zero-delay, the block length $n$ equals to $1$ and the encoding function is a sample-by-sample one. In the next section, we introduce our zero-delay JSCC encoding function and two kinds of corresponding decoding schemes.

%when $\Delta$ is small, the summation in (\ref{power}) approximately equals to the area under the curve representing the probability density function of $S_i$, which is $1$. 
%The decoding function at the $i$-th receiver, $\phi^n_i: \mathbb{R}^n\rightarrow \mathbb{R}^n$, for $i=1,2$, estimates $S^n_i$, i.e., $\hat{S}^n_i=\phi^n_i(Y^n_i)$.

\section{Proposed Schemes}
Our proposed HDA encoder is depicted in Fig.~\ref{figr:framework}. The digital information $T_i=Q(S_i)$ is produced by a midtread scalar quantizer with reconstruction levels $\{t_k\}_{k=-\infty}^{\infty}, k\in\mathbb{Z}$, where $\mathbb{Z}$ denotes the set of integers. We use $\Delta=t_k-t_{k-1}$ to denote quantization step. Then it holds that $t_k = k\Delta$. In parallel, the analog part is used to send the source itself. The scaled combination of $T_i$ and $S_i$, $X_i = f_i(S_i) =  \delta_i T_i+\beta_iS_i$, is then transmitted through the channel. The average transmit power from encoder $i$ is
\begin{align}
 %P_i&=\mathbb{E}[X_i^2]\nonumber\\
    P_i  &=\mathbb{E}\left[\left(\delta_iT_i+\beta_iS_i\right)^2\right]\nonumber\\
      &=\delta_i^2 \mathbb{E}[T_i^2]+\beta_i^2+2\delta_i\beta_i\mathbb{E}[T_iS_i] \quad\quad i=1,2,
\label{equ1}
\end{align}
where
\begin{align}
\mathbb{E}[T_iS_i]&= \sum\limits_{k}\int\limits_{t_k-\frac{\Delta}{2}}^{t_k+\frac{\Delta}{2}}t_ksP_{S_i}(s_i)ds_i\nonumber\\
&= \sum\limits_{k}\frac{t_k}{\sqrt{2\pi}}\times\Big[\exp\big(-\frac{\big(t_k-\frac{\Delta}{2}\big)^2}{2}\big)-\exp\big(-\frac{\big(t_k+\frac{\Delta}{2}\big)^2}{2}\big)\Big]\nonumber\\
&=\sum\limits_{k}\frac{\Delta}{\sqrt{2\pi}}\times \exp\Big(-\frac{\big(t_k-\frac{\Delta}{2}\big)^2}{2}\Big),
\label{11}
\end{align}
\begin{align}
%\mathbb{E}[T_i^2]&=\sum\limits_{k}t^2_kp_{T_i}(t_k)\nonumber\\
%\mathbb{E}[T_i^2]&=\sum\limits_{k}t^2_k\int\limits_{t_k-\frac{\Delta}{2}}^{t_k+\frac{\Delta}{2}}P_{S_i}(s_i)ds_i\nonumber\\
\mathbb{E}[T_i^2]&=\sum\limits_{k}t^2_k\times \frac{1}{2}\Big[\erf\big(\frac{t_k+\frac{\Delta}{2}}{\sqrt{2}}\big)-\erf\big(\frac{t_k-\frac{\Delta}{2}}{\sqrt{2}}\big)\Big].
\label{12}
\end{align}

As shown in Fig.~\ref{figr:framework}, at $i$-th receiver, $(T_i, T_{i^c})$ are firstly recovered by a joint estimator, then $S_i$ is reconstructed by utilizing the analog channel output $Y_i$ and the estimated digital information pair $(\hat{T}_i,\hat{T}_{i^c})$ jointly. We propose two kinds of decoding schemes and our object is to find minimum \emph{average} distortion $D$ with the \emph{average} power constraint $P_1+P_2=2P$.

\subsection{Scheme A}
%For the ease of analysis, we limit $k$, the number of quantization levels,% by $k_{\min}$ which satisfies $k_{\min}-\frac{1}{2}\Delta=-\kappa$, where $Pr(S_i\in[-\kappa, \kappa])\approx 1$, that is, the overload distortion approximately equals to $0$.  
Given the correlation $\rho$, we set the maximum distance between the quantization output of $S_i$ and the one of $S_{i^c}$ as
 \[
M\Delta=\bigg\lceil\frac{3\sqrt{1-\rho^2}+\Big[\big(|k|_{\max}-\frac{1}{2}\big)\Delta-\rho\big(|k|_{\max}-\frac{1}{2}\big)\Delta\Big]}{\Delta}\bigg\rceil\times\Delta.
\]
The derivation of $M\Delta$ can be found in Appendix A.
 $k$ denotes integer quantization index. Its absolute value is limited by $|k|_{\max}$, which satisfies $(|k|_{\max}+\frac{1}{2})\Delta=\kappa$, where $Pr(S_i\in[-\kappa, \kappa])\approx 1$, that is, the overload distortion approximately equals to $0$. 

We apply maximum a posterior (MAP) estimator for recovery of digital information,
\begin{align*}
 (\hat{t}_i, \hat{t}_{i^c})=\arg\!\!\!\!\max_{\substack{t_{i,k},t_{i^c,k'}\\t_{i^c,k'}\in[t_{i,k}-M\Delta,t_{i,k}+M\Delta]}} \!\!\!\!\!\!\!\!\!\!P_{(T_i,T_{i^c}),Y_i}\Big(\big(t_{i,k},t_{i^c,k'}\big),y_i\Big).
 \end{align*}
Herein, 
\begin{align*}
&\lefteqn{P_{(T_i,T_{i^c}),Y_i}\Big(\big(t_{i,k},t_{i^c,k^'}\big),y_i\Big)}\\
&=\int^{\infty}_{-\infty}\int^{\infty}_{-\infty}P_{S_i, S_{i^c},(T_i,T_{i^c}),Y_i}\Big(s_i, s_{i^c}, (t_{i,k},t_{i^c,k^'}), y_i\Big)ds_ids_{i^c}\\
&=\int^{t_{i,k}+\frac{\Delta}{2}}_{t_{i,k}-\frac{\Delta}{2}}\int^{t_{i^c,k^'}+\frac{\Delta}{2}}_{t_{i^c,k^'}-\frac{\Delta}{2}}P_{S_i,S_{i^c}}(s_i,s_{i^c})P_{W_i}(u_{t_{i,k},t_{i^c,k^'}})ds_ids_{i^c}, 
\end{align*}
where
\begin{align*}
u_{t_{i,k},t_{i^c,k^'}}=y_i-\delta_i t_{i,k}-\beta_is_i-c_{i^c}(\delta_{i^c} t_{i^c,k^'}+\beta_{i^c}s_{i^c}).
\end{align*}
After obtaining $(\hat{t}_i, \hat{t}_{i^c})$, $\hat{S}_i$ is estimated as below,
 \begin{align*}
\hat{s}_i  &= \mathbb{E}[S_i|(T_{i},T_{i^c}),Y_i]\\
      &=\int\limits_{-\infty}^{\infty}s_iP_{S_i|T_i,T_{i^c},Y_i}\Big(s_i|\big(\hat{t}_i,\hat{t}_{i^c}\big),y_i\Big)ds_i\\
      %&=&\frac{\int\limits_{-\infty}^{\infty}s_iP_{S_i, (T_i,T_{i^c}),Y}(s_i, (\hat{t}_i, \hat{t}_{i^c}),y_i)ds_i}{\int\limits_{-\infty}^{\infty}P_{S_i, (T_i,T_{i^c}),Y}(s_i, (\hat{t}_i, \hat{t}_{i^c}),y_i)ds_i}\\
      &=  \frac{\int\limits_{-\infty}^{\infty}\int\limits_{-\infty}^{\infty}s_iP_{S_i,S_{i^c},(T_i,T_{i^c}),Y_i}(s_i,s_{i^c}, \hat{t}_i, \hat{t}_{i^c},y_i)ds_ids_{i^c}}{\int\limits_{-\infty}^{\infty}\int\limits_{-\infty}^{\infty}P_{S_i, S_{i^c},(T_i,T_{i^c}),Y_i}(s_i,s_{i^c},\hat{t}_i, \hat{t}_{i^c},y_i)ds_ids_{i^c}}\\
      &=\frac{\int^{\hat{t}_i+\frac{\Delta}{2}}_{\hat{t}_i-\frac{\Delta}{2}}\int^{\hat{t}_{i^c}+\frac{\Delta}{2}}_{\hat{t}_{i^c}-\frac{\Delta}{2}}s_iP_{S_i,S_{i^c}}(s_i,s_{i^c})P_{W_i}(u_{\hat{t}_i, \hat{t}_{i^c}})ds_ids_{i^c}}{\int^{\hat{t}_i+\frac{\Delta}{2}}_{\hat{t}_i-\frac{\Delta}{2}}\int^{\hat{t}_{i^c}+\frac{\Delta}{2}}_{\hat{t}_{i^c}-\frac{\Delta}{2}}P_{S_i,S_{i^c}}(s_i,s_{i^c})P_{W_i}(u_{\hat{t}_i, \hat{t}_{i^c}})ds_ids_{i^c}}.
 \end{align*}
For Scheme A, it is hard to obtain analytical form for $D$. 
 \subsection{Scheme B}
 Note that $X_i$ can be rewritten as the summation of quantized value and quantization error $R_i=S_i-T_i$,
  \[
  X_i = \delta_iT_i+\beta_i(T_i+R_i)=\alpha_iT_i+\beta_iR_i,
  \]
 where $\alpha_i$ denotes $\delta_i+\beta_i$.
 As $R_i$ is constrained in $[-\frac{\Delta}{2}, \frac{\Delta}{2}]$, we propose a pseudo maximum likelihood (ML) estimator as follows,
\begin{align*}
(\hat{t}_i, \hat{t}_{i^c}) = \arg\!\!\!\!\min_{\substack{t_{i,k},t_{i^c,k'}\\t_{i^c,k'}\in[t_{i,k}-M\Delta,t_{i,k}+M\Delta]}} \!\!\!\!\!\!\!\!y_i-\alpha_it_{i,k}-c_{i^c}\alpha_{i^c}t_{i^c,k'}.
\end{align*}
	
 As
 \begin{align*}
 Y_i &= \alpha_iT_i+\beta_iR_i+c_{i^c}\bigg[\delta_{i^c}T_{i^c}+\beta_{i^c}\Big(\rho\big(T_i+R_i\big)+N\Big)\bigg]+W_i\\
       & =(\alpha_i+c_{i^c}\beta_{i^c}\rho)T_i+c_{i^c}(\alpha_{i^c}-\beta_{i^c})T_{i^c}\\
       &+(\beta_i+c_{i^c}\beta_{i^c}\rho)R_i+c_{i^c}\beta_{i^c}N+W_i,
 \end{align*}
the quantization error $R_i$ is estimated linearly by 
 \[
 \hat{r}_i = \Gamma_i\Big[y_i-\big(\alpha_i+c_{i^c}\beta_{i^c}\rho\big)\hat{t}_i-c_{i^c}\big(\alpha_{i^c}-\beta_{i^c}\big)\hat{t}_{i^c}\Big],
 \] 
 where $\Gamma_i$ is linear coefficient.
Finally, $\hat{S}_i = \hat{T}_i+\hat{R}_i$.

\textbf{Distortion Analysis:} 
 For Scheme B, we can express the overall distortion $D_i$ in analytical form. By definition,
   \begin{align}
   &D_i= \mathbb{E}[(T_i+R_i-\hat{T}_i-\hat{R}_i)^2]\nonumber\\
   &=\mathbb{E}\Bigg[\bigg(\lambda+\Big(1-\Gamma_i\big(\beta_i+c_{i^c}\beta_{i^c}\rho\big)\Big)R_i-\Gamma_i c_{i^c}\beta_{i^c}N-\Gamma_i W_i\bigg)^2\Bigg]\nonumber\\
   &=\mathbb{E}[\lambda^2]+\Big(1-\Gamma_i\big(\beta_i+c_{i^c}\beta_{i^c}\rho\big)\Big)^2\sigma_R^2+(\Gamma_i c_{i^c}\beta_{i^c})^2\sigma_N^2+\Gamma_i^2\sigma_W^2\nonumber\\
   &+2\Big(1-\Gamma_i\big(\beta_i+c_{i^c}\beta_{i^c}\rho\big)\Big)\Big(1-\Gamma_i\big(\alpha_i+c_{i^c}\beta_{i^c}\rho\big)\Big)\mathbb{E}\Big[R_i\big(T_i-\hat{T}_i\big)\Big]\nonumber\\
   &-2\Big(1-\Gamma_i\big(\beta_i+c_{i^c}\beta_{i^c}\rho\big)\Big)\Gamma_i c_{i^c}(\alpha_{i^c}-\beta_{i^c})\mathbb{E}\Big[R_i\big(T_{i^c}-\hat{T}_{i^c}\big)\Big],
   \label{equ2}
   \end{align}
     where
   \[
   \lambda=\Big(1-\Gamma_i\big(\alpha_i+c_{i^c}\beta_{i^c}\rho\big)\Big)(T_i-\hat{T}_i)-\Gamma_i c_{i^c}(\alpha_{i^c}-\beta_{i^c})(T_{i^c}-\hat{T}_{i^c}).
   \]
    The justification of (\ref{equ2}) can be found in Appendix B.
    
  Next, we would analyze the components of (\ref{equ2}) one by one. 
   \begin{align}
  \mathbb{E}[\lambda^2]=\sum_k\sum\limits_{m=k-M}^{k+M}\sum\limits_l\sum\limits_{n=l-M}^{l+M} \Phi P_{\hat{T}_{i^c},\hat{T}_i,T_{i^c},T_i}(t_n,t_l,t_m,t_k),
   \label{equ3}
   \end{align}
   where
  \begin{align*}
   \Phi=\Big((1-\Gamma_i\big(\alpha_i+c_{i^c}\beta_{i^c}\rho\big))(t_k-t_l)-\Gamma_i c_{i^c}\big(\alpha_{i^c}-\beta_{i^c}\big)\big(t_m-t_n\big)\Big)^2.
   \end{align*}
\begin{align}
\mathbb{E}[R_iT_i]&=\mathbb{E}[T_iS_i]-\mathbb{E}[T_i^2].
\label{equ4}
 \end{align} 
$\mathbb{E}[R_iT_i]$ can be obtained by substituting (\ref{11}) and (\ref{12}) into (\ref{equ4}).  
 \begin{align}
\mathbb{E}[R_iT_{i^c}]&=\mathbb{E}\Big[\mathbb{E}\big[R_iT_{i^c}|T_i\big]\Big]\nonumber\\
&=\sum_kP_{T_i}(t_k)\mathbb{E}[R_iT_{i^c}|T_i = t_k]\nonumber\\
&=\sum_kP_{T_i}(t_k)\sum\limits_m\int\limits^{\frac{\Delta}{2}}_{\frac{-\Delta}{2}}t_{m}r_iP_{R_i,T_{i^c}|T_i}(r_i,t_m|t_k)dr_i\nonumber\\
 &=\sum_k\sum\limits_m\int\limits^{\frac{\Delta}{2}}_{\frac{-\Delta}{2}}t_mr_iP_{R_i,T_{i^c},T_i}(r_i,t_m,t_k)dr_i\nonumber\\
 %&=\sum_k\sum\limits_l\int^{\Delta/2}_{-\Delta/2}rt_lp_{T^c_i|R_i,T_i}(t_l|r,t_k)p_S(t_k+r)dr\\
% &=\sum_k\sum\limits_l\int^{\Delta/2}_{-\Delta/2}rt_lp_{T^c_i,T_i,R_i}(t_l,t_k,r)dr\\
 %&=\sum_k\int^{\Delta/2}_{-\Delta/2}\int^{\Delta/2}_{-\Delta/2}rt_mp_{T_{i^c},R_{i^c},T_i, R_i}(t_l,r^c,t_k,r)dr^cdr\nonumber\\
  &=\sum_k\sum\limits_{m=k-M}^{k+M}t_m\int\limits^{t_k+\frac{\Delta}{2}}_{t_k-\frac{\Delta}{2}}\int\limits^{t_m+\frac{\Delta}{2}}_{t_m-\frac{\Delta}{2}}(s_i-t_k)P_{S_{i^c},S_i}(s_{i^c},s_i)ds_{i^c}ds_i.
  \label{equ5}
 \end{align}    
 The derivation of (\ref{equ5}) can be found in Appendix D. 
 \begin{align}
\mathbb{E}[R_i\hat{T}_i]&=\mathbb{E}\Big[\mathbb{E}\big[R_i\hat{T_i}|T_i\big]\Big]\nonumber\\
 %&=\sum_kp_{T_i}(t_k)\mathbb{E}[R_i\hat{T_i}|T_i = t_k]\nonumber\\
 %&=\sum_kp_{T_i}(t_k)\sum\limits_l\int\limits^{\frac{\Delta}{2}}_{\frac{-\Delta}{2}}t_{l}r_ip_{R_i,\hat{T}_i|T_i}(r,t_l|t_k)dr_i\nonumber\\
 &=\sum_k\sum\limits_l\int\limits^{\frac{\Delta}{2}}_{\frac{-\Delta}{2}}t_lr_iP_{R_i,\hat{T}_i,T_i}(r_i,t_l,t_k)dr_i\nonumber\\
 &=\sum_k\sum\limits_{m=k-M}^{k+M}\sum\limits_l\sum\limits_{n=l-M}^{l+M}t_l\int\limits^{\frac{\Delta}{2}}_{\frac{-\Delta}{2}}\int\limits^{\frac{\Delta}{2}}_{\frac{-\Delta}{2}}r_i\Psi dr_{i^c}dr_i,
 \label{equ6}
 \end{align}
 where
 \[
 \Psi =  P_{\hat{T}_{i^c},\hat{T}_i,T_{i^c},T_i,R_{i^c},R_i}(t_n,t_l,t_m,t_k,r_{i^c},r_i).
 \]
 
 Similarly, 
  \begin{align}
\mathbb{E}[R_i\hat{T}_{i^c}]&=\mathbb{E}\Big[\mathbb{E}\big[R_i\hat{T}_{i^c}|T_i\big]\Big]\nonumber\\
 %&=\sum_k\sum\limits_l\int^{\Delta/2}_{-\Delta/2}rt_lp_{\hat{T}^c_i|R_i,T_i}(t_l|r,t_k)p_S(t_k+r)dr\\
 &=\sum_k\sum\limits_{m=k-M}^{k+M}\sum\limits_l\sum\limits_{n=l-M}^{l+M}t_n\int\limits^{\frac{\Delta}{2}}_{\frac{-\Delta}{2}}\int\limits^{\frac{\Delta}{2}}_{\frac{-\Delta}{2}}r_i\Psi dr_{i^c}dr_i.
 \label{equ7}
 \end{align} 

 To calculate (\ref{equ3}), (\ref{equ6}), (\ref{equ7}), we need to find the value of joint probability $P_{\hat{T}_{i^c},\hat{T}_i,T_{i^c},T_i}(t_n,t_l,t_m,t_k)$. 

Given the digital information pair $(T_i,T_{i^c})=(t_k,t_m)$, the distance $d$ between the recovered digital message $\alpha_it_l+c_{i^c}\alpha_{i^c}t_n$ and the original digital message $\alpha_it_k+c_{i^c}\alpha_{i^c}t_m$ must be one of the values in the set \footnote{The size of the set can be shrunk through limiting the absolute values of the elements by $2\times\bigg(5\sigma_W+\big(\beta_i+c_{i^c}\beta_{i^c}\big)\frac{\Delta}{2}\bigg)$.} as below,
\begin{align*}	
\{&(p\alpha_i+qc_{i^c}\alpha_{i^c})\Delta\}, \\
\textrm{where}\quad&p, q\in\mathbb{Z}, m+q\in[k+p-M,k+p+M]
%\textrm{where}\quad& p\in\{0,1, ..., 2M-1\}, \\
%&q\geq0, |p-q|\in\{0,1, ..., M\}, 
\end{align*}

%$2$%\times(3\sigma_W+(\beta_i+\beta_{i^c})\frac{\Delta}{2})$.
Sort the set in ascending order and check the position of $d$ in it. We can obtain the one just before $d$ and the one right after $d$. These two distances are denoted respectively by $d_l$ and $d_u$. Then $P_{\hat{T}_{i^c},\hat{T}_i,T_i,T_i}(t_n,t_l,t_m,t_k)$ can be expressed as follows,
  \begin{align}
 \lefteqn{P_{\hat{T}_{i^c},\hat{T}_i,T_{i^c},T_i}(t_n,t_l,t_m,t_k)}\nonumber\\
%&=\int\limits^{\frac{\Delta}{2}}_{\frac{-\Delta}{2}}\int\limits^{\frac{\Delta}{2}}_{\frac{-\Delta}{2}}P_{\hat{T}_{i^c},\hat{T}_i,T_{i^c},T_i,R_{i^c},R_i}(t_n,t_l,t_m,t_k,r_{i^c},r_i)dr_{i^c}dr_i\nonumber\\
&=\int\limits^{t_k+\frac{\Delta}{2}}_{t_k-\frac{\Delta}{2}}\int\limits^{t_m+\frac{\Delta}{2}}_{t_m-\frac{\Delta}{2}}P_{\hat{T}_{i^c},\hat{T}_i,S_{i^c},S_i}(t_n,t_l,s_{i^c},s_i)ds_{i^c}ds_i\nonumber\\
%&=\int\limits^{t_k+\frac{\Delta}{2}}_{t_k-\frac{\Delta}{2}}\int\limits^{t_m+\frac{\Delta}{2}}_{t_m-\frac{\Delta}{2}}p_{W_i}(d_l+\frac{d-d_l}{2}\leq w_i+\mu\leq d+\frac{d_u-d}{2}-\mu)\\
%&p_{S_{i^c},S_i}(s_{i^c},s_i)ds_{i^c}ds_i\\
  &=\int\limits^{t_k+\frac{\Delta}{2}}_{t_k-\frac{\Delta}{2}}\int\limits^{t_m+\frac{\Delta}{2}}_{t_m-\frac{\Delta}{2}}\int\limits^{ub}_{lb}P_{W_i}(w_i)P_{S_{i^c},S_i}(s_{i^c},s_i)dw_ids_{i^c}ds_i,
  \label{eqtnjoint}
  \end{align}
where $ub = \frac{d+d_u}{2}-\mu, lb = \frac{d+d_l}{2}-\mu$,
and $\mu = \beta_i(s_i-t_k)+c_{i^c}\beta_{i^c}(s_{i^c}-t_m)$.
 The derivation of (\ref{eqtnjoint}) can be found in Appendix C.
%\begin{equation}
%ub= \left\{
%\begin{array}
%&  d>0\\
   %          d_u+\frac{d-d_u}{2}-\mu,& d<0
 %\end{array}
%\right.
%\end{equation}
% $$ lb =\left\{
%\begin{aligned}
% d_l+\frac{d-d_l}{2}-\mu,  \quad\quad\quad d>0
 %            d+\frac{d_l-d}{2}-\mu   \quad\quad\quad d<0
%\end{aligned}
%\right.
%$$

%If $d=0$, which means that no recovery errors occur, the probability is calculated by $1-\sum\limits_{d\neq0}P_{\hat{T}_{i^c},\hat{T}_i,T_i,T_i}(t_n,t_l,t_m,t_k)$.
   %Take a combination of parameters for example. When $\Delta = 0.4, \alpha_1 = 0.4, \alpha_2 = 2.2, \rho = 0.95, M = 3, c_1 = c_2 =0.8$, the procedure to find the probability that the digital message pair $(-0.4, 0.4)$ is recovered incorrectly as $(0,0)$ at decoder 1 is as follows. Firstly, the distance $d$ in this case is $|\alpha_1\times0.4-c2\times\alpha_2\times0.4|$, i.e., $0.7520$. Then we check the position of $d$ in the ed distances set. The one prior to $d$ is $d_l=0.6240$, and the successor of $d$ is $d_u=0.88$. Finally, the probability required is calculated by
%\begin{align*}
%\lefteqn{P_{\hat{T}_{i^c},\hat{T},T_{i^c},T_i}(0,0,0.4,-0.4)} \\
%&= \int\limits_{-0.6}^{-0.2}\int\limits_{0.2}^{0.6}\int\limits_{0.6240+0.064-\mu}^{0.752+0.064-\mu}P_{W_i}(w_i)P_{S_{i^c},S_i}(s_{i^c},s_i)dw_ids_{i^c}ds_i. 
%\end{align*}

 According to (\ref{eqtnjoint}),
 \begin{align}
\int\limits^{\frac{\Delta}{2}}_{\frac{-\Delta}{2}}\int\limits^{\frac{\Delta}{2}}_{\frac{-\Delta}{2}}r_i\Psi dr_{i^c}dr_i=\int\limits^{t_k+\frac{\Delta}{2}}_{t_k-\frac{\Delta}{2}}\int\limits^{t_m+\frac{\Delta}{2}}_{t_m-\frac{\Delta}{2}}\int\limits^{ub}_{lb}(s_i-t_k)P_{W_i}(w_i)P_{S_{i^c},S_i}(s_{i^c},s_i)dw_ids_{i^c}ds_i \label{equ8}.
 \end{align}
 The derivation of (\ref{equ8}) can be found in Appendix D. 
Substituting (\ref{eqtnjoint}) into (\ref{equ3}), and (\ref{equ8}) into (\ref{equ6}), (\ref{equ7}), we obtain all the components needed to calculate (\ref{equ2}). 

 \begin{figure*}[ht]
  \begin{tabular}{cc} 
  \hspace{-0.5cm}
 \subfigure[]{ 
    \begin{minipage}[t]{0.33\textwidth} 
    \centering 
    \includegraphics[height = 4.5cm,width = 6cm]{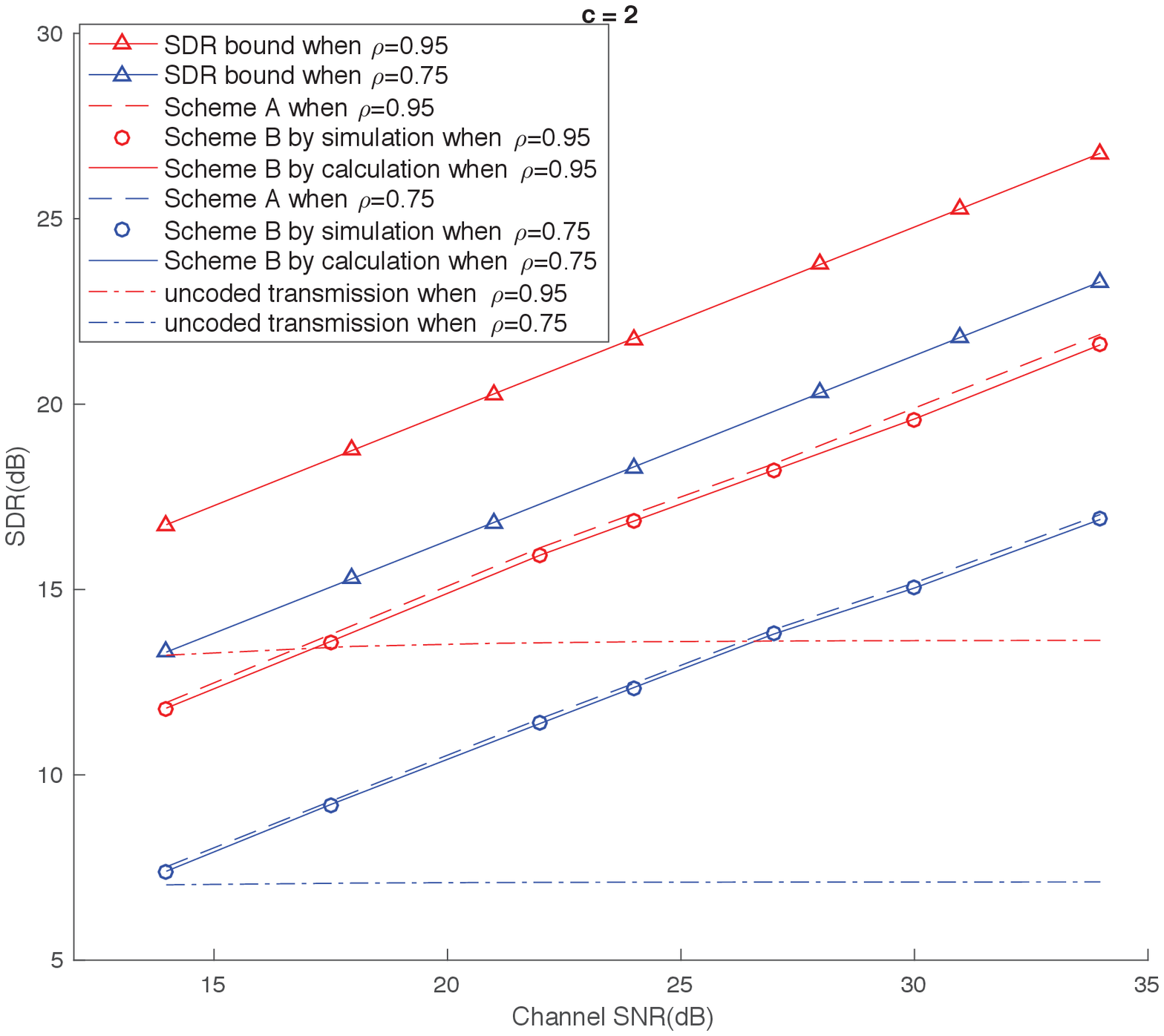} 
    \end{minipage}} 
    \hspace{-0.3cm}
    \subfigure[]{ 
    \begin{minipage}[t]{0.33\textwidth} 
    \centering 
    \includegraphics[height = 4.5cm,width = 6cm]{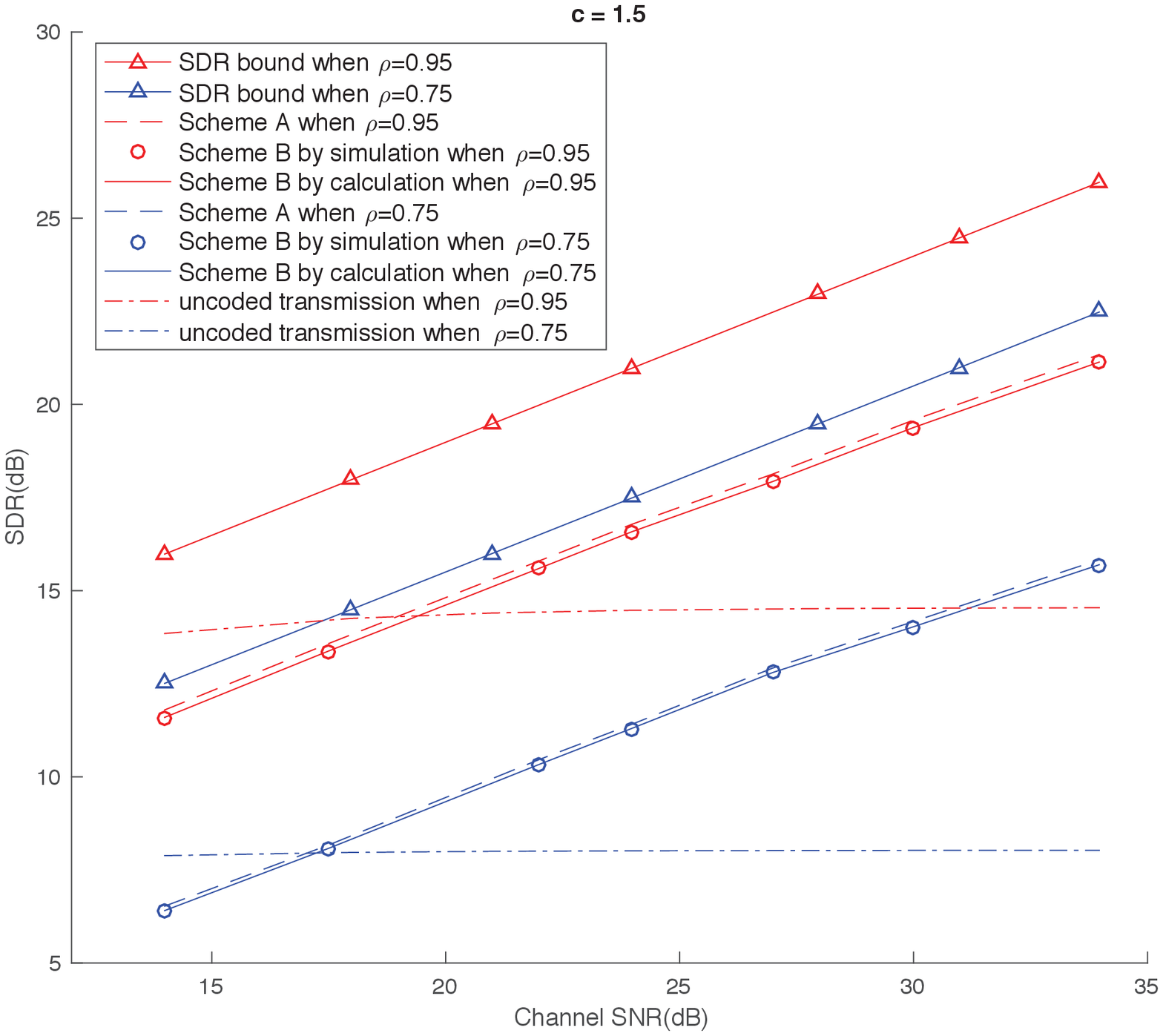} 
    \end{minipage}} 
     \hspace{-0.3cm}
    %\subfigure[]{ 
    %\begin{minipage}[t]{0.33\textwidth} 
    %\centering 
    %\includegraphics[height = 5cm,width = 6cm]{figurec12} 
    %\end{minipage}} 
    \subfigure[]{ 
    \begin{minipage}[t]{0.33\textwidth} 
    \centering 
    \includegraphics[height = 4.5cm,width = 6cm]{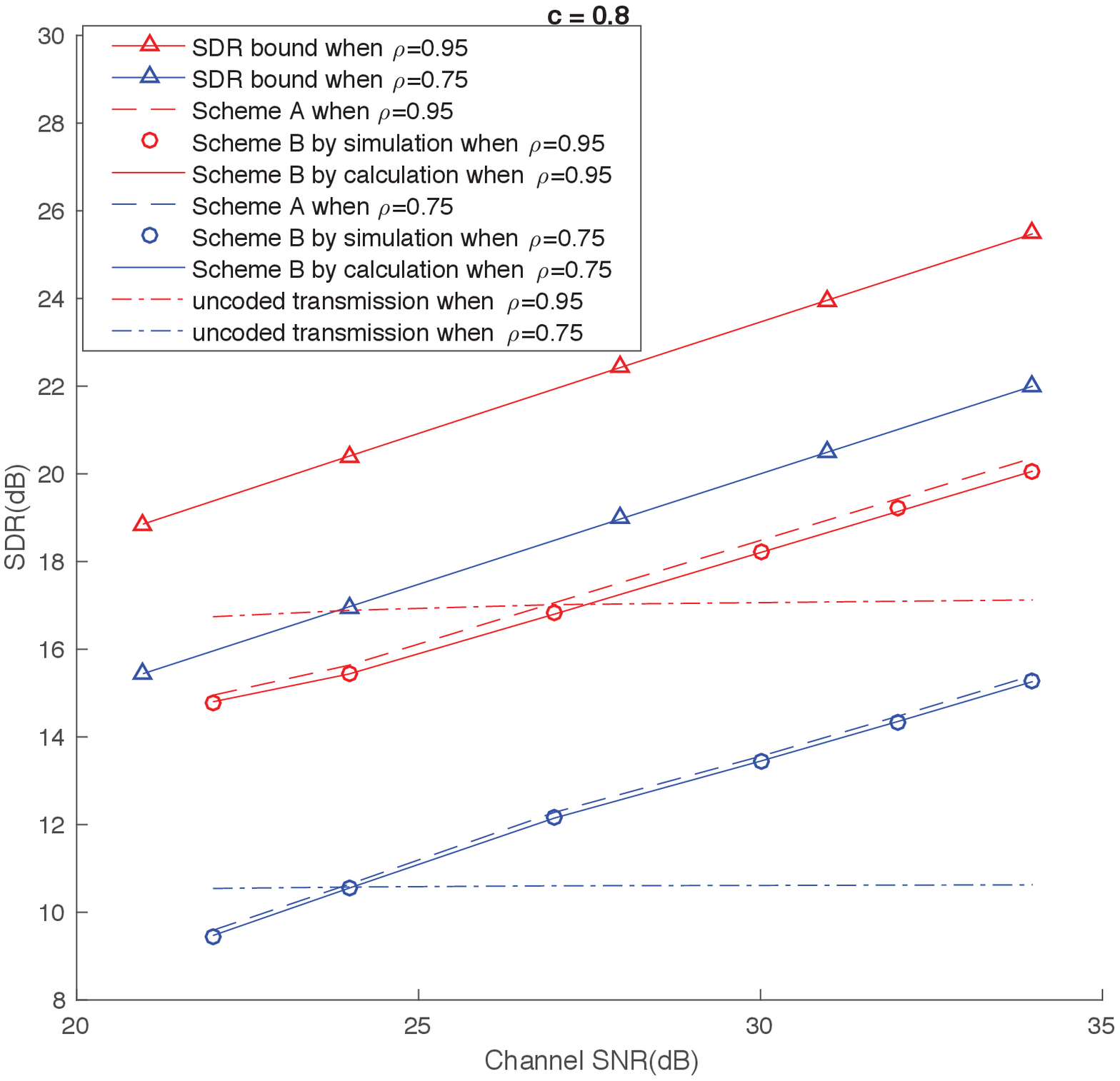} 
    \end{minipage}} 
    \end{tabular}
\begin{tabular}{cc} 
  \hspace{-0.5cm}
 \subfigure[]{ 
    \begin{minipage}[t]{0.33\textwidth} 
    \centering 
    \includegraphics[height = 4.5cm,width = 6cm]{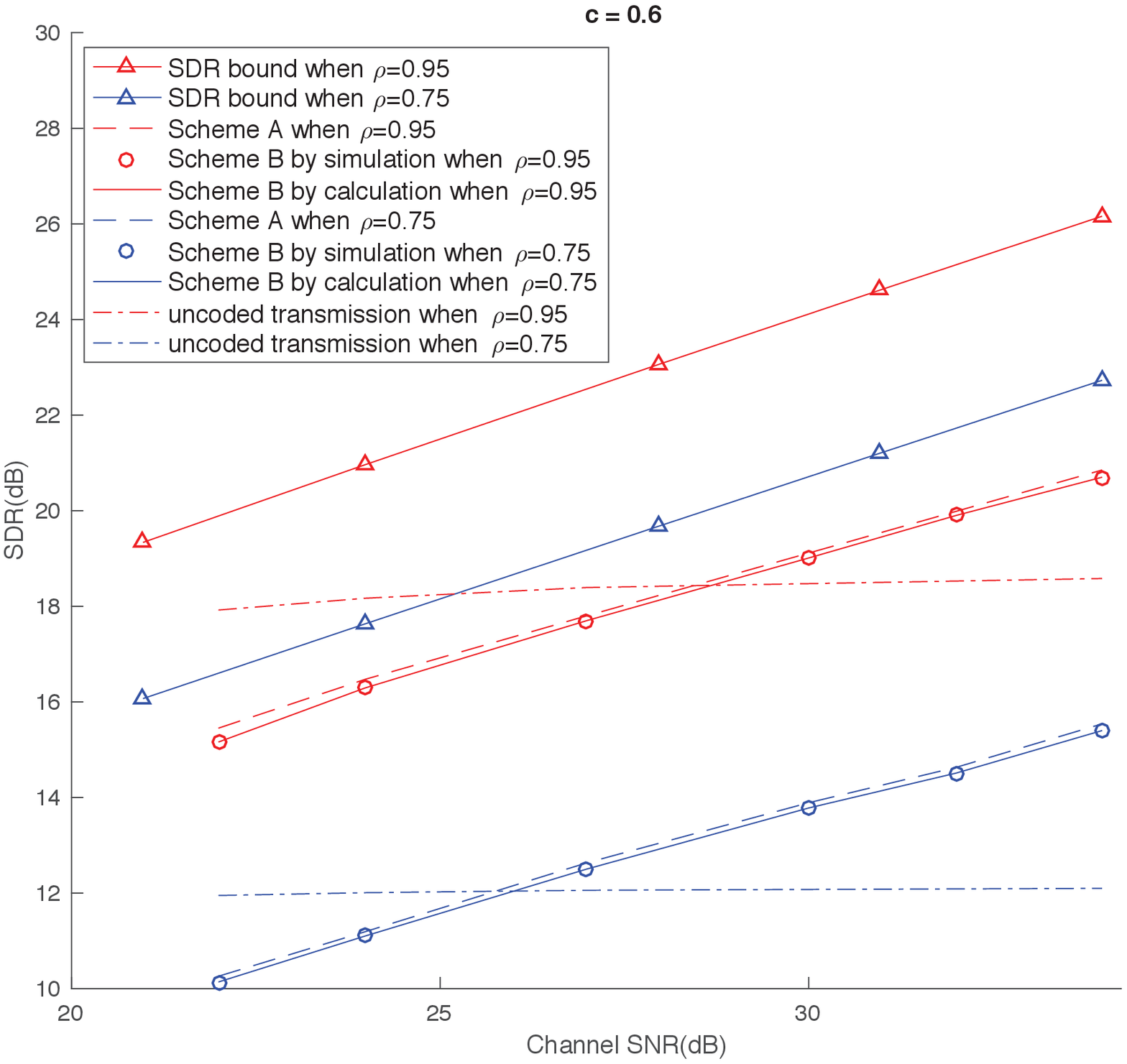} 
    \end{minipage}} 
    \hspace{-0.3cm}
    \subfigure[]{ 
    \begin{minipage}[t]{0.33\textwidth} 
    \centering 
    \includegraphics[height = 4.5cm,width = 6cm]{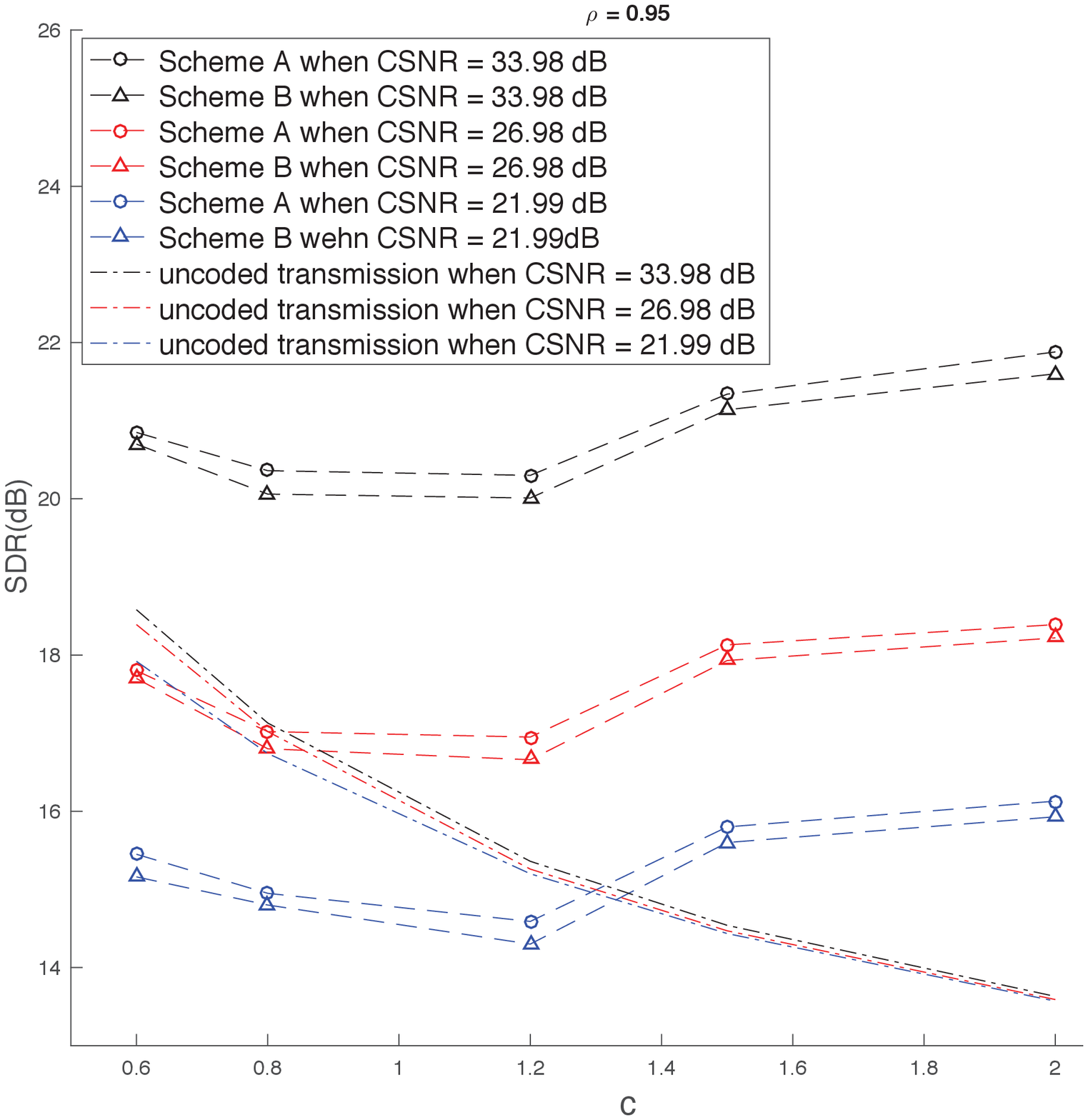} 
    \end{minipage}} 
     \hspace{-0.3cm}
    %\subfigure[]{ 
    %\begin{minipage}[t]{0.33\textwidth} 
    %\centering 
    %\includegraphics[height = 5cm,width = 6cm]{figurec12} 
    %\end{minipage}} 
    \subfigure[]{ 
    \begin{minipage}[t]{0.33\textwidth} 
    \centering 
    \includegraphics[height = 4.5cm,width = 6cm]{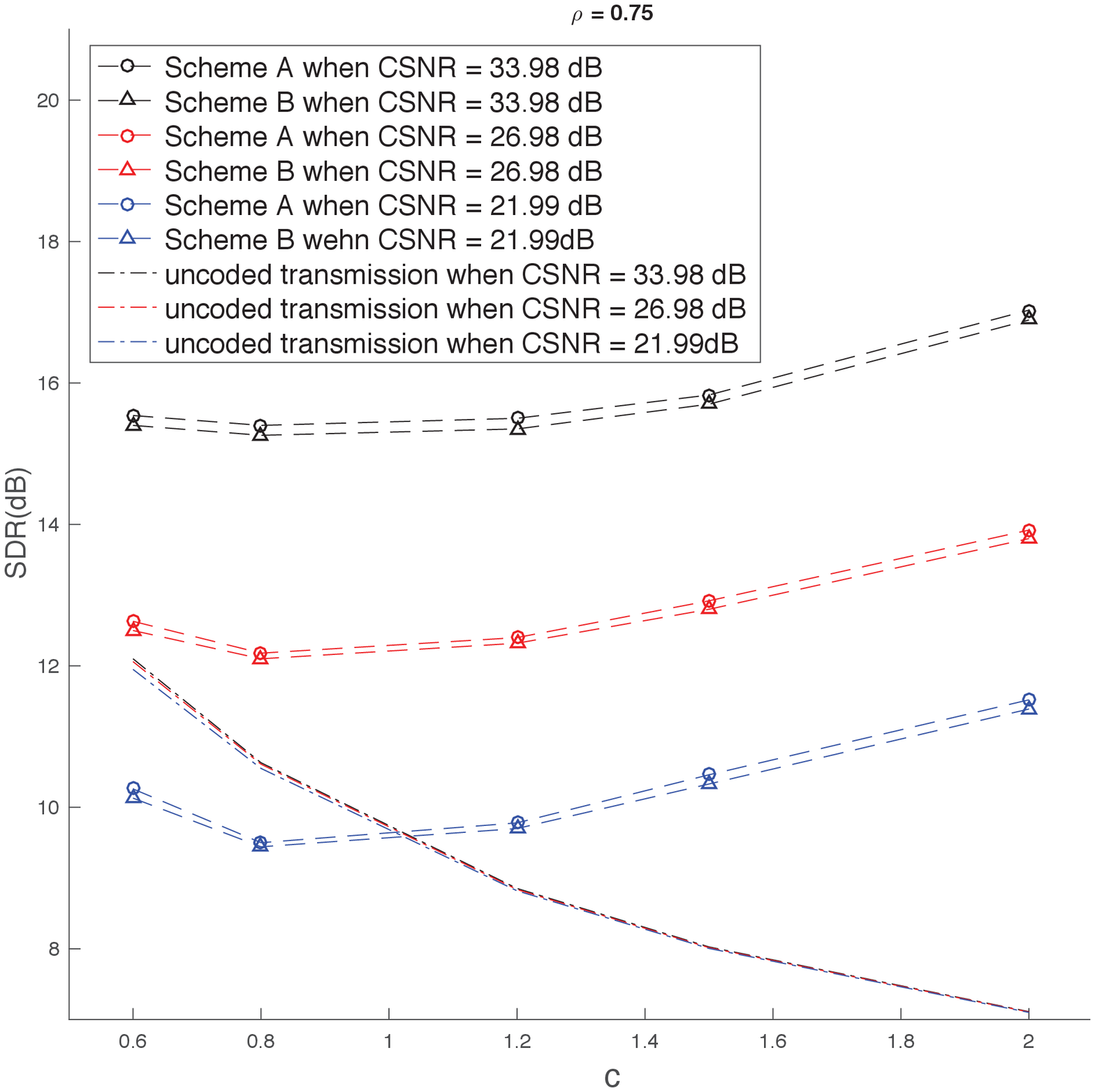} 
    \end{minipage}} 
    \end{tabular}
\caption{(a)-(d) Performance comparisons of relevant schemes, (e)-(f) Performance trends of relevant schemes with varying $c$.}
\label{fig2} 
\end{figure*}

\section{simulation results}
We compare our schemes with uncoded transmission and upper bounds. Though the schemes are proposed for general case, here we only present the symmetric interference case, that is, $c_1=c_2=c$. The performance is measured by signal-to-distortion ratio $SDR = \frac{\sigma_1^2}{D}$. In our experiments, $|k|_{\max}$, the maximum value of $|k|$ is chosen to be $|k|_{\max} = \lceil\frac{6}{\Delta}-\frac{1}{2}\rceil$. $|k|$ denotes the absolute value of integer quantization index $k$. Fig.~\ref{fig2} (a), (b) show the comparison results when the interference is strong. Herein, the upper bound is obtained by applying the lower bound resulting from Lemma 3 in \cite{denial2}. The results while the interference is weak can be found in Fig.~\ref{fig2} (c), (d). The upper bound here is derived through Lemma 1 in \cite{denial1}. The markers 'o' represent the simulation results by Scheme B using optimal parameters obtained from minimizing the \emph{average} end-to-end distortion after (\ref{equ2}) is substituted, while the lines without marker show the calculation results by Scheme B. The effectiveness of (\ref{equ2}) is demonstrated as markers $'o'$ basically stick to corresponding lines. From all these figures, it can be told that both Scheme A and Scheme B we proposed are superior to the uncoded transmission when CSNR is larger than a threshold value for different interference values and correlation coefficients. Fixing interference factor $c$, the superiority of our schemes towards uncoded transmission becomes more obvious when $\rho$ decreases,  which is reflected by the fact that the threshold value of CSNR gets smaller.
 
Comparing the Fig.~\ref{fig2} (a), (b) with the figures (c), (d), the advantage of our schemes on uncoded transmission is more prominent for the cases with strong interference. In the scenarios with weak interference, the advantage decays with the decreasing of $c$. This is reasonable as when $c\rightarrow 0$, the interference channel degrades to point to point channel, and as well known, uncoded transmission achieves optimal distortion in such scenario. In Fig.~\ref{fig2} (e), (f), we fix the correlation $\rho$ and plot $SDR$ as a function of $c$ with varying CSNR. Note that the performances of proposed schemes enhance with the increasing of $c$ in the presence of strong interference and with the decreasing of $c$ for weak interference channel while the performance of uncoded transmission always deteriorates with the increasing of $c$. 
%Fig.~\ref{fig2}.(f) shows the optimal parameters obtained from the equation (\ref{equ2}) while scheme B is applied when $c=2, CSNR = 33.9794dB$. When correlation $\rho$ decreases, the quantization step $\Delta$ increases,  then more analog information is superimposed, which is reflected on the fact that $\beta$ increases. %There exists gap between $\alpha_1$ and $\alpha_2$ to make sure that the distances among different digital message $\alpha_1t_{1,k}+\alpha_2t_{2,k'}$ are large enough so that digital information pair can be recovered correctly with high probability.
\begin{appendices}
\section{Derivation of $M\Delta$}
As illustrated in the Section II, $(S_1,S_2)$ are correlated gaussian sources with zero mean and covariance matrix
\[
\left[
\begin{array}{cc}
1 & \rho \\
\rho & 1
\end{array}
\right] .\; 
\]

Then 
\begin{align*}
\mathbb{E}[S_{i^c}|S_i=s_i] &= \mathbb{E}[S_{i^c}]+\rho(\frac{\sigma_{S_{i^c}}}{\sigma_{S_i}})(s_i-\mathbb{E}[S_i])\\
                                             &= \rho s_i.\\
                                             \mathbb{VAR}[S_{i^c}|S_i] &= \sigma^2_{S_{i^c}}(1-\rho^2)\\
                                                                        & = (1-\rho^2).
\end{align*}
In other words, $S_{i^c}|(S_i=s_i) \sim{\mathcal N}(\rho s_i,1-\rho^2)$. Then with probability $p$, where $p\approx 1$, $S_{i^c}$ falls into the interval $[\rho s_i-3\sqrt{1-\rho^2}, \rho s_i+3\sqrt{1-\rho^2}]$. \footnote{We have tried $4\sqrt{1-\rho^2}$ and $5\sqrt{1-\rho^2}$ and there were no advanced performance improvements.}

For a midtread quantizer, $k\Delta$ denotes reconstruction levels while $(k+\frac{1}{2})\Delta$ denotes decision levels.

\begin{figure}[ht]
\centering
\includegraphics[scale=0.6]{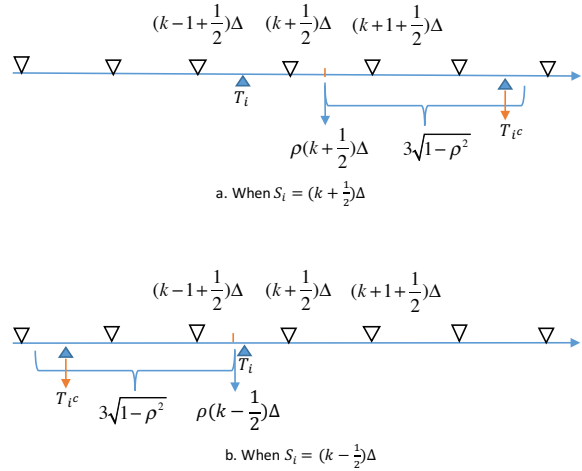}
\caption{Illustration of maximum distance}
\label{figr:ill}
\end{figure}

In the following, we assume that $k<0$. Fig.~\ref{figr:ill} illustrates how the maximum distance is calculated when $k<0$.  $T_i = k\Delta$ supposing that $S_i$ falls in the interval $[(k-\frac{1}{2})\Delta, (k+\frac{1}{2})\Delta]$. Fig.~\ref{figr:ill}(a) shows the maximum distance between $T_i$ and $T_{i^c}$ if $S_i$ equals to the right boundary value $(k+\frac{1}{2})\Delta$ while (b) shows the maximum distance when $S_i$ equals to the left boundary value $(k-\frac{1}{2})\Delta$. Obviously, the maximum distance in (a) is larger than the one in (b). As a result, we only need to analyze the expression of the maximum distance in Fig.~\ref{figr:ill} (a). As depicted in (a), 
\begin{align}
d_{\max}(k) &= M\Delta \nonumber\\
               &= \bigg\lceil\frac{3\sqrt{1-\rho^2}+\Big[\rho\big(k+\frac{1}{2}\big)\Delta-\big(k+\frac{1}{2}\big)\Delta\Big]}{\Delta}\bigg\rceil\times\Delta \nonumber\\
               &= \bigg\lceil\frac{3\sqrt{1-\rho^2}-\Big[\big(k+\frac{1}{2}\big)\Delta-\rho\big(k+\frac{1}{2}\big)\Delta\Big]}{\Delta}\bigg\rceil\times\Delta
               \label{eqt}
\end{align}
The procedure to find $d_{\max(k)}$ for the case $k>=0$ is similar due to the symmetry. However, the expression becomes 
\[
d_{\max}(k) = \bigg\lceil\frac{3\sqrt{1-\rho^2}+\Big[\big(k-\frac{1}{2}\big)\Delta-\rho\big(k-\frac{1}{2}\big)\Delta\Big]}{\Delta}\bigg\rceil\times\Delta.
\]
%Therefore, we emphasized that $k<0$ in Section III since we adopted (\ref{eqt}) to express maximum distance in the paper. 

So $d_{\max}(k)$ is a increasing function of absolute value of $k$ as below
\[
d_{\max}(|k|) = \bigg\lceil\frac{3\sqrt{1-\rho^2}+\Big[\big(|k|-\frac{1}{2}\big)\Delta-\rho\big(|k|-\frac{1}{2}\big)\Delta\Big]}{\Delta}\bigg\rceil\times\Delta.
\]
 If $|k|$ goes to infinity, $d_{\max}$ would go to infinity as well. As matter of fact, the absolute value of $k$ is limited. We limite $|k|$ by $|k|_{\max}$ which satisfies $(|k|_{\max}\Delta+\frac{1}{2}\Delta)=\kappa$, where $Pr(S_i\in[-\kappa, \kappa])\approx 1$, that is, the overload distortion approximately equals to $0$. 
 As a result, the maximum distance is set to be
 \[
M\Delta=\bigg\lceil\frac{3\sqrt{1-\rho^2}+\Big[\big(|k|_{\max}-\frac{1}{2}\big)\Delta-\rho\big(|k|_{\max}-\frac{1}{2}\big)\Delta\Big]}{\Delta}\bigg\rceil\times\Delta.
\]
as shown in the paper.
 In our experiments, we choose $|k|_{\max} = \lceil\frac{6}{\Delta}-\frac{1}{2}\rceil$ so that $\kappa > 6$. 

\section{Justification of (\ref{equ2})}
  % D_i&= \mathbb{E}[(T_i+R_i-\hat{T}_i-\hat{R}_i)^2]\nonumber\\
  %&=\mathbb{E}\[(T_i+R_i-\hat{T}_i-\Gamma_i\big(Y_i-(\alpha_i+c_{i^c}\beta_{i^c}\rho)\hat{T}_i\big)]\\
   %&=\mathbb{E}\Bigg[\bigg(\lambda+\Big(1-\Gamma_i\big(\beta_i+c_{i^c}\beta_{i^c}\rho\big)\Big)R_i-\Gamma_i c_{i^c}\beta_{i^c}N-\Gamma_i W_i\bigg)^2\Bigg]\nonumber\\
  % &=\mathbb{E}[\lambda^2]+\Big(1-\Gamma_i\big(\beta_i+c_{i^c}\beta_{i^c}\rho\big)\Big)^2\sigma_R^2+(\Gamma_i c_{i^c}\beta_{i^c})^2\sigma_N^2+\Gamma_i^2\sigma_W^2\nonumber\\
  % &+2\Big(1-\Gamma_i\big(\beta_i+c_{i^c}\beta_{i^c}\rho\big)\Big)\Big(1-\Gamma_i\big(\alpha_i+c_{i^c}\beta_{i^c}\rho\big)\Big)\mathbb{E}\Big[R_i\big(T_i-\hat{T}_i\big)\Big]\nonumber\\
  
   %&-2\Big(1-\Gamma_i\big(\beta_i+c_{i^c}\beta_{i^c}\rho\big)\Big)\Gamma_i c_{i^c}(\alpha_{i^c}-\beta_{i^c})\mathbb{E}\Big[R_i\big(T_{i^c}-\hat{T}_{i^c}\big)\Big],

  % \end{align}

Substituting
\[
 \hat{R}_i = \Gamma_i\Big[Y_i-\big(\alpha_i+c_{i^c}\beta_{i^c}\rho\big)\hat{T}_i-c_{i^c}\big(\alpha_{i^c}-\beta_{i^c}\big)\hat{T}_{i^c}\Big]
 \] 
 into 
 \[
 D_i\nonumber  = \mathbb{E}[(T_i+R_i-\hat{T}_i-\hat{R}_i)^2],\\
 \] 
 we have
   \begin{align}
   &D_i\nonumber\\
   %&= \mathbb{E}[(T_i+R_i-\hat{T}_i-\hat{R}_i)^2]\nonumber\\
   & =\mathbb{E}\Bigg[\bigg(T_i+R_i-\hat{T}_i-\Gamma_i\Big(Y_i-\big(\alpha_i+c_{i^c}\beta_{i^c}\rho\big)\hat{T}_i-c_{i^c}(\alpha_{i^c}-\beta_{i^c})\hat{T}_{i^c}\Big)\bigg)^2\Bigg]\nonumber\\
    &=\mathbb{E}\Bigg[\bigg(T_i+R_i-\hat{T}_i-\Gamma_i\Big(\alpha_iT_i+\beta_iR_i+c_{i^c}\alpha_{i^c}T_{i^c}+c_{i^c}\beta_{i^c}(\rho(T_i+R_i)+N-T_{i^c})+W_i-\psi\Big)\bigg)^2\Bigg]\label{1}\\
    &=\mathbb{E}\Bigg[\bigg((1-\Gamma_i(\alpha_i+c_{i^c}\beta_{i^c}\rho))(T_i-\hat{T}_i)-\Gamma_i c_{i^c}(\alpha_{i^c}-\beta_{i^c})(T_{i^c}-\hat{T}_{i^c})\nonumber\\
    &+\Big(1-\Gamma_i\big(\beta_i+c_{i^c}\beta_{i^c}\rho\big)\Big)R_i-\Gamma_i c_{i^c}\beta_{i^c}N-\Gamma_i W_i\bigg)^2\Bigg]\nonumber\\
    &=\mathbb{E}\Bigg[\bigg(\lambda+\Big(1-\Gamma_i\big(\beta_i+c_{i^c}\beta_{i^c}\rho\big)\Big)R_i-\Gamma_i c_{i^c}\beta_{i^c}N-\Gamma_i W_i\bigg)^2\Bigg]\nonumber\\
   &=\mathbb{E}[\lambda^2]+\Big(1-\Gamma_i\big(\beta_i+c_{i^c}\beta_{i^c}\rho\big)\Big)^2\sigma_R^2+(\Gamma_i c_{i^c}\beta_{i^c})^2\sigma_N^2+\Gamma_i^2\sigma_W^2\nonumber\\
   &+2\Big(1-\Gamma_i\big(\beta_i+c_{i^c}\beta_{i^c}\rho\big)\Big)\Big(1-\Gamma_i\big(\alpha_i+c_{i^c}\beta_{i^c}\rho\big)\Big)\mathbb{E}\Big[R_i\big(T_i-\hat{T}_i\big)\Big]\nonumber\\
   &-2\Big(1-\Gamma_i\big(\beta_i+c_{i^c}\beta_{i^c}\rho\big)\Big)\Gamma_i c_{i^c}(\alpha_{i^c}-\beta_{i^c})\mathbb{E}\Big[R_i\big(T_{i^c}-\hat{T}_{i^c}\big)\Big]\nonumber,
   \end{align}

  where
   \[
   \lambda=\Big(1-\Gamma_i\big(\alpha_i+c_{i^c}\beta_{i^c}\rho\big)\Big)(T_i-\hat{T}_i)-\Gamma_i c_{i^c}(\alpha_{i^c}-\beta_{i^c})(T_{i^c}-\hat{T}_{i^c}).
   \]
   In (\ref{1}), we use $\psi$ to denote $\big(\alpha_i+c_{i^c}\beta_{i^c}\rho\big)\hat{T}_i+c_{i^c}(\alpha_{i^c}-\beta_{i^c})\hat{T}_{i^c}$ due to the space limit.

\section{Derivation of (\ref{eqtnjoint})}
Suppose that $(T_i,T_{i^c})=(t_k,t_m), (\hat{T}_i,\hat{T}_{i^c})=(t_l,t_n)$, then by the description of the midtread quantizer in the first paragraph of Section III, $(T_i,T_{i^c})=(k\Delta,m\Delta), (\hat{T}_i,\hat{T}_{i^c})=(l\Delta,n\Delta)$. As a result, $\alpha_i(t_l-t_k)+c_{i^c}\alpha_{i^c}(t_n-t_m)$, which is distance $d$ between the recovered digital message $\alpha_it_l+c_{i^c}\alpha_{i^c}t_n$ and the original digital message $\alpha_it_k+c_{i^c}\alpha_{i^c}t_m$ , equals to 
\begin{align*}
\alpha_i(l-k)\Delta+c_{i^c}\alpha_{i^c}(n-m)\Delta.
\end{align*}
According to the definition of pseudo ML estimator,
\begin{align}
(\hat{t}_i, \hat{t}_{i^c}) = \arg\!\!\!\!\min_{\substack{t_{i,k},t_{i^c,k'}\\t_{i^c,k'}\in[t_{i,k}-M\Delta,t_{i,k}+M\Delta]}} \!\!\!\!\!\!\!\!y_i-\alpha_it_{i,k}-c_{i^c}\alpha_{i^c}t_{i^c,k'},
\label{q}
\end{align}
$t_n$ should be in the interval $[t_l-M\Delta,t_l+M\Delta]$. In other words, $n\in[l-M, l+M]$. Assuming that $
l-k = p, n-m = q$ and $p,q\in\mathbb{Z}$, consequently, $m+q\in[k+p-M,k+p+M]$. 

\begin{figure}[ht]
\centering
\includegraphics[width = 8cm, height = 5cm]{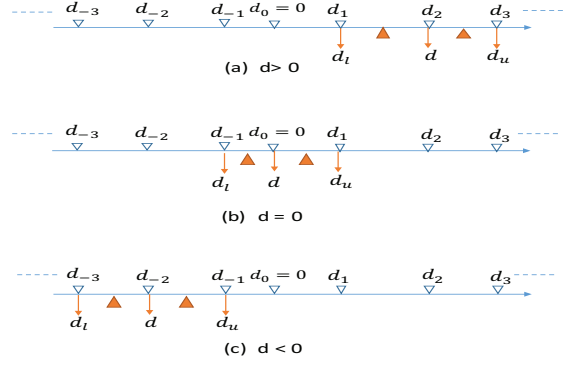}
\caption{Illustration on how to obtain the upper bound and lower bound for three cases.}
\label{figr2}
\end{figure}

Let us sort all the possible distances and label them on the real axis, as shown in Fig.~\ref{figr2}\footnote{Though the number of possible distances can be infinite, we shrink the size by limiting $d$ by $\abs{d}\leq 2\times(5\sigma_W+\frac{\Delta}{2}(\beta_i+c_i\beta_{i^c}))$}. The origin $d_{0}$ denotes the case that $d=0$, i.e., $T_i, T_{i^c}$ are correctly recovered. $d_{i}, i>0$ denotes the i-th distance that is larger than $0$ while $d_{i}, i<0$ denotes the i-th one that is smaller than $0$. 

When would the event that $S_i=s_i, (s_i\in[t_k-\frac{\Delta}{2},t_k+\frac{\Delta}{2}]$), $S_{i^c}=s_{i^c}, (s_{i^c}\in[t_m-\frac{\Delta}{2},t_m+\frac{\Delta}{2}]), \hat{T}_{i^c}=t_n,\hat{T}_i=t_l$ happen? Let us make an example for illustration. If $d=d_2$, then $d_l = d_1, d_u = d_3$. According to (\ref{q}), the event occurs when $\alpha_it_k+c_{i^c}\alpha_{i^c}t_m+\beta_i(s_i-t_k)+c_{i^c}\beta_{i^c}(s_{i^c}-t_m)+w_i-\alpha_it_{l}-c_{i^c}\alpha_{i^c}t_n$ is minimum, i.e., $\beta_i(s_i-t_k)+c_{i^c}\beta_{i^c}(s_{i^c}-t_m)+w_i-d_1$ is smaller than the difference between $\beta_i(s_i-t_k)+c_{i^c}\beta_{i^c}(s_{i^c}-t_m)+w_i$ and all other $d$ . In other words, $\beta_i(s_i-t_k)+c_{i^c}\beta_{i^c}(s_{i^c}-t_m)+w_i$ falls in the interval [$d_1+\frac{d_2-d_1}{2}, d_2+\frac{d_3-d_2}{2}$]. It concludes that 
 \begin{align*}
 \lefteqn{P_{\hat{T}_{i^c},\hat{T}_i,T_{i^c},T_i}(t_n,t_l,t_m,t_k)}\nonumber\\
&=\int\limits^{\frac{\Delta}{2}}_{\frac{-\Delta}{2}}\int\limits^{\frac{\Delta}{2}}_{\frac{-\Delta}{2}}P_{\hat{T}_{i^c},\hat{T}_i,T_{i^c},T_i,R_{i^c},R_i}(t_n,t_l,t_m,t_k,r_{i^c},r_i)dr_{i^c}dr_i\nonumber\\
&=\int\limits^{t_k+\frac{\Delta}{2}}_{t_k-\frac{\Delta}{2}}\int\limits^{t_m+\frac{\Delta}{2}}_{t_m-\frac{\Delta}{2}}P_{\hat{T}_{i^c},\hat{T}_i,S_{i^c},S_i}(t_n,t_l,s_{i^c},s_i)ds_{i^c}ds_i\\
&=\int\limits^{t_k+\frac{\Delta}{2}}_{t_k-\frac{\Delta}{2}}\int\limits^{t_m+\frac{\Delta}{2}}_{t_m-\frac{\Delta}{2}}p_{W_i}\Big(d_1+\frac{d_2-d_1}{2}\leq w_i+\beta_i\big(s_i-t_k\big)+c_{i^c}\beta_{i^c}\big(s_{i^c}-t_m\big)\leq d_2+\frac{d_3-d_2}{2}\Big)p_{S_{i^c},S_i}(s_{i^c},s_i)ds_{i^c}ds_i\\
 &=\int\limits^{t_k+\frac{\Delta}{2}}_{t_k-\frac{\Delta}{2}}\int\limits^{t_m+\frac{\Delta}{2}}_{t_m-\frac{\Delta}{2}}\int\limits^{ub}_{lb}P_{W_i}(w_i)P_{S_{i^c},S_i}(s_{i^c},s_i)dw_ids_{i^c}ds_i,
  \label{eqtnjoint}
  \end{align*}
  where $ub = d_2+\frac{d_3-d_2}{2}-\Big(\beta_i\big(s_i-t_k\big)+c_{i^c}\beta_{i^c}\big(s_{i^c}-t_m\big)\Big)$, $lb = d_1+\frac{d_2-d_1}{2}-\Big(\beta_i\big(s_i-t_k)+c_{i^c}\beta_{i^c}(s_{i^c}-t_m)\Big)$.
  
 The same procedure applies for the cases $d=0$ and $d<0$. Fig.~\ref{figr2} (a), (b), (c) depict the three conditions respectively. As a result,
 \begin{equation*}
ub = \left\{
             \begin{array}{lcl}
             d+\frac{d_u-d}{2}-\mu, &\text{if} &d>0 \\
             d_u+\frac{d-d_u}{2}-\mu,&\text{if} &d<0\\
             \frac{d_u}{2}-\mu&\text{if} &d=0
             \end{array}  
        \right.
\end {equation*}
\begin{equation*}
lb = \left\{
             \begin{array}{lcl}
            d_l+\frac{d-d_l}{2}-\mu, &\text{if} &d>0 \\
             d+\frac{d_l-d}{2}-\mu &\text{if} &d<0\\
             \frac{d_l}{2}-\mu&\text{if} &d=0
             \end{array}  
        \right.
\end {equation*}
 
The expressions of $ub$ and $lb$ can be unified into one form instead of three individual forms for three cases. So we modified the expression by 
 \[ub = \frac{d+d_u}{2}-\mu, lb = \frac{d+d_l}{2}-\mu\]

\section{Derivations of (\ref{equ5}) and (\ref{equ8})}
\textbf{Derivations of (\ref{equ5}):}
 \begin{align*}
&\mathbb{E}[R_iT_{i^c}]=\mathbb{E}\Big[\mathbb{E}\big[R_iT_{i^c}|T_i\big]\Big]\nonumber\\
&=\sum_kP_{T_i}(t_k)\mathbb{E}[R_iT_{i^c}|T_i = t_k]\nonumber\\
&=\sum_kP_{T_i}(t_k)\sum\limits_m\int\limits^{\frac{\Delta}{2}}_{\frac{-\Delta}{2}}t_{m}r_iP_{R_i,T_{i^c}|T_i}(r,t_m|t_k)dr_i\nonumber\\
 &=\sum_k\sum\limits_m\int\limits^{\frac{\Delta}{2}}_{\frac{-\Delta}{2}}t_mr_iP_{R_i,T_{i^c},T_i}(r_i,t_m,t_k)dr_i\nonumber\\
 &=\sum_k\sum\limits_m\int\limits^{\frac{\Delta}{2}}_{\frac{-\Delta}{2}}\int\limits^{\frac{\Delta}{2}}_{\frac{-\Delta}{2}}t_mr_iP_{R_{i^c},R_i,T_{i^c},T_i}(r_{i^c},r_i,t_m,t_k)dr_{i^c}dr_i\nonumber\\
   &\overset{a}=\sum_k\sum\limits_{m=k-M}^{k+M}t_m\int\limits^{t_k+\frac{\Delta}{2}}_{t_k-\frac{\Delta}{2}}\int\limits^{t_m+\frac{\Delta}{2}}_{t_m-\frac{\Delta}{2}}(s_i-t_k)P_{S_{i^c},S_i}(s_{i^c},s_i)ds_{i^c}ds_i,
 \end{align*}    
 where (a) follows because
  \begin{align*}
P_{R_{i^c},R_i,T_{i^c},T_i}(r_{i^c},r_i,t_m,t_k)&=P_{S_{i^c},S_i,T_{i^c},T_i}(r_{i^c}+t_m,r_i+t_k,t_m,t_k)\\
&=P_{S_{i^c},S_i}(r_{i^c}+t_m,r_i+t_k)P_{T_{i^c},T_i|S_{i^c},S_i}(t_m,t_k|r_{i^c}+t_m,r_i+t_k)\\
&=P_{S_{i^c},S_i}(r_{i^c}+t_m,r_i+t_k)
 \end{align*}
 
\textbf{Derivations of (\ref{equ8}):}
 \begin{align*}
\lefteqn{\int\limits^{\frac{\Delta}{2}}_{\frac{-\Delta}{2}}\int\limits^{\frac{\Delta}{2}}_{\frac{-\Delta}{2}}r_iP_{\hat{T}_{i^c},\hat{T}_i,T^c_i,T_i,R_{i^c},R_i}(t_n,t_l,t_m,t_k,r_{i^c},r_i) dr_{i^c}dr_i}\nonumber\\
&\overset{b}=\int\limits^{t_k+\frac{\Delta}{2}}_{t_k-\frac{\Delta}{2}}\int\limits^{t_m+\frac{\Delta}{2}}_{t_m-\frac{\Delta}{2}}(s_i-t_k)P_{\hat{T}_{i^c},\hat{T}_i,S_{i^c},S_i}(t_n,t_l,s_{i^c},s_i)ds_{i^c}ds_i\nonumber\\
  &=\int\limits^{t_k+\frac{\Delta}{2}}_{t_k-\frac{\Delta}{2}}\int\limits^{t_m+\frac{\Delta}{2}}_{t_m-\frac{\Delta}{2}}\int\limits^{ub}_{lb}(s_i-t_k)P_{W_i}(w_i)P_{S_{i^c},S_i}(s_{i^c},s_i)dw_ids_{i^c}ds_i,
   \end{align*}
where (b) follows from the same fact with (a).
\end{appendices}
 
\end{document}